\documentclass[journal]{IEEEtran}
\IEEEoverridecommandlockouts

\usepackage{xfrac}
\usepackage{amsmath}
\usepackage{multirow,etoolbox}
\usepackage{color}
\usepackage{amssymb}
\usepackage{cite}
\usepackage{mathrsfs}
\usepackage{diagbox}

\usepackage{subcaption}
\usepackage[utf8]{inputenc}
 
\newcommand{\mathbbm}[1]{\text{\usefont{U}{bbm}{m}{n}#1}} \usepackage{colortbl}
\usepackage{stfloats}
\usepackage{relsize}

\usepackage[subtle]{savetrees}

\newcommand{\Sum}{\mathlarger{\sum}}

\makeatletter
\patchcmd{\@maketitle}
  {\addvspace{0.5\baselineskip}\egroup}
  {\addvspace{-1\baselineskip}\egroup}
  {}
  {}
\makeatother

\begin{document}
\bstctlcite{IEEEexample:BSTcontrol}

\title{\makebox[\linewidth]{\parbox{\dimexpr\textwidth+1.5cm\relax}{\centering Error Performance Analysis of Multi-user Detection in Uplink-NOMA with  Adaptive $\mathcal{M}$-QAM}}}
\author{Hichem Semira, Ferdi Kara,~\IEEEmembership{Senior Member,~IEEE,} Hakan Kaya, Halim Yanikomeroglu,~\IEEEmembership{Fellow,~IEEE}. 
\thanks{The work of F. Kara is supported by TÜBİTAK under 2219 Postdoctoral Scholarship. H. Semira is with Laboratory of Electronics and New Technologies (LENT), University of Oum El Bouaghi, Oum El Bouaghi, Algeria, email: hichem.semira@univ-oeb.dz. F. Kara is with Department of Computer Engineering, Zonguldak Bulent Ecevit University (ZBEU), Turkey and is also with the Department of Systems and Computer Engineering (SCE), Carleton University (CU), Ottawa, K1S 5B6, ON, Canada, e-mail:  e-mail: f.kara@beun.edu.tr.  H. Kaya is with the Department of Electrical-Electronics Engineering, ZBEU, Turkey  e-mail:  hakan.kaya @beun.edu.tr. H. Yanikomeroglu is with the Department of SCE, CU, Ottawa, K1S 5B6, ON, Canada, e-mail: halim@sce.carleton.ca.}}
\maketitle
\begin{abstract}
This work provides a generalized performance analysis for the multi-user uplink-NOMA system with adaptive square quadrature amplitude modulation (QAM) over Rayleigh fading channels. Motivated by the massive IoT connections and unavailability of orthogonal resources for each node, we consider a multi-access scheme where multi-users with single-antenna transmit data to a multiple-antenna base station through the same resource block. By taking advantage of combining diversity paths with the proposed joint maximum-likelihood detector (JMLD), a closed form expression for the upper bound of bit error rate (BER) is obtained. Despite the number of users or the order of modulation, the analytical results endorsed via computer simulations reveal the ability of the MRC-JMLD detector to discard the error floor completely. Moreover, the simulation results show that the MRC-JMLD surpasses its counterparts significantly and ensures a full diversity order.     
\end{abstract}


\IEEEpeerreviewmaketitle
\section{Introduction}

Non-Orthogonal Multiple Access (NOMA) scheme is considered as auspicious candidate to achieve the massive connection and improved spectral efficiency in ultra dense networks (e.g., Internet-of-Things (IoT) applications \cite{Yuan2021}) due to meet some of the challenges regarding the shortage of radio spectrum. However, this comes at the cost of  excessive multi-access interference. Typically, to eliminate interference at the receiver, the successive interference cancellation detector (SICD) has been widely considered as the preferred detector to separate the superimposed data of users. 

In terms of information-theoretic limits, the SICD may seem an optimal solution and may reach the potential of NOMA by theoretically eliminating the interference in both downlink and uplink \cite{Ding2020}. However, by considering the practical implementation of the SICD, despite the effectiveness in downlink scenario \cite{Kara2018d}, it suffers considerably from the presence of the error floor in high SNR regime for the uplink scenario \cite{Kara2018d,Kara2020}. This is due to the imperfect elimination of interference which propagates from an iteration to the next. Recently, the joint maximum likelihood detector (JMLD) has attracted growing interest in both downlink and uplink schemes \cite{assaf21, yahya21, Yeom2019,Shahab2020, Shahab2021,Semira2021,semira2021_wcl,He2021}. Even though the JMLD processing to detect the users’ signal is different from that of SICD, their performance is identical in downlink scheme either under perfect or imperfect channel state information (CSI) \cite{assaf21}. This made the derivation of an exact BER expression possible and simpler in a downlink scheme for a larger number of users served by different modulation order \cite{yahya21}. By virtue of performance matching, SICD is mostly preferred in the downlink scheme. On the other hand, in uplink scenario, the JMLD has proved its superiority and effectiveness by surmounting the error floor problem encountered in SICD \cite{Yeom2019,Shahab2020, Shahab2021,Semira2021,semira2021_wcl,He2021}.
However, aforementioned studies are mostly based on only computer simulations \cite{Shahab2020,Shahab2021,He2021}  and  are limited to very special cases \cite{Yeom2019,Semira2021,semira2021_wcl}. Theoretical bit error rate (BER) analysis is presented in a two-users scenario for QPSK \cite{Yeom2019} and $\mathcal{M}$-PSK \cite{Semira2021} whereas the analysis is extended for multi-user with $\mathcal{M}$-PSK in \cite{semira2021_wcl}. However, to the best of the authors' knowledge, there has been no study yet to present a generalized error performance of JMLD with arbitrary $\mathcal{M}$-QAM. Considering the facts that NOMA has potential for multi-user scenarios (more than two users) to enable massive connectivity required systems (e.g., IoT \cite{Yuan2021}) and the recent 3GPP standards implement $\mathcal{M}$-QAM in the physical layer \cite{3gpp1630}, the evaluation of JMLD with $\mathcal{M}$-QAM for multi-user scenarios is still an open research problem. To this end, in this paper, we analyze the error performance of JMLD in multi-user uplink NOMA with an adaptive $\mathcal{M}$-QAM, where the modulation order is chosen according to the channel quality indicator (CQI) as being in standards \cite{3gpp1630}. We derive a very tight closed-form upper bound for the BER expression and validate the result via computer simulations. We also show the reliability of the JMLD by outperforming its counterparts and removing the error floor while meeting the recommendations of the standards \cite{3gpp1630}.
\begin{figure}
    \centering
    \includegraphics[width=0.6\columnwidth]{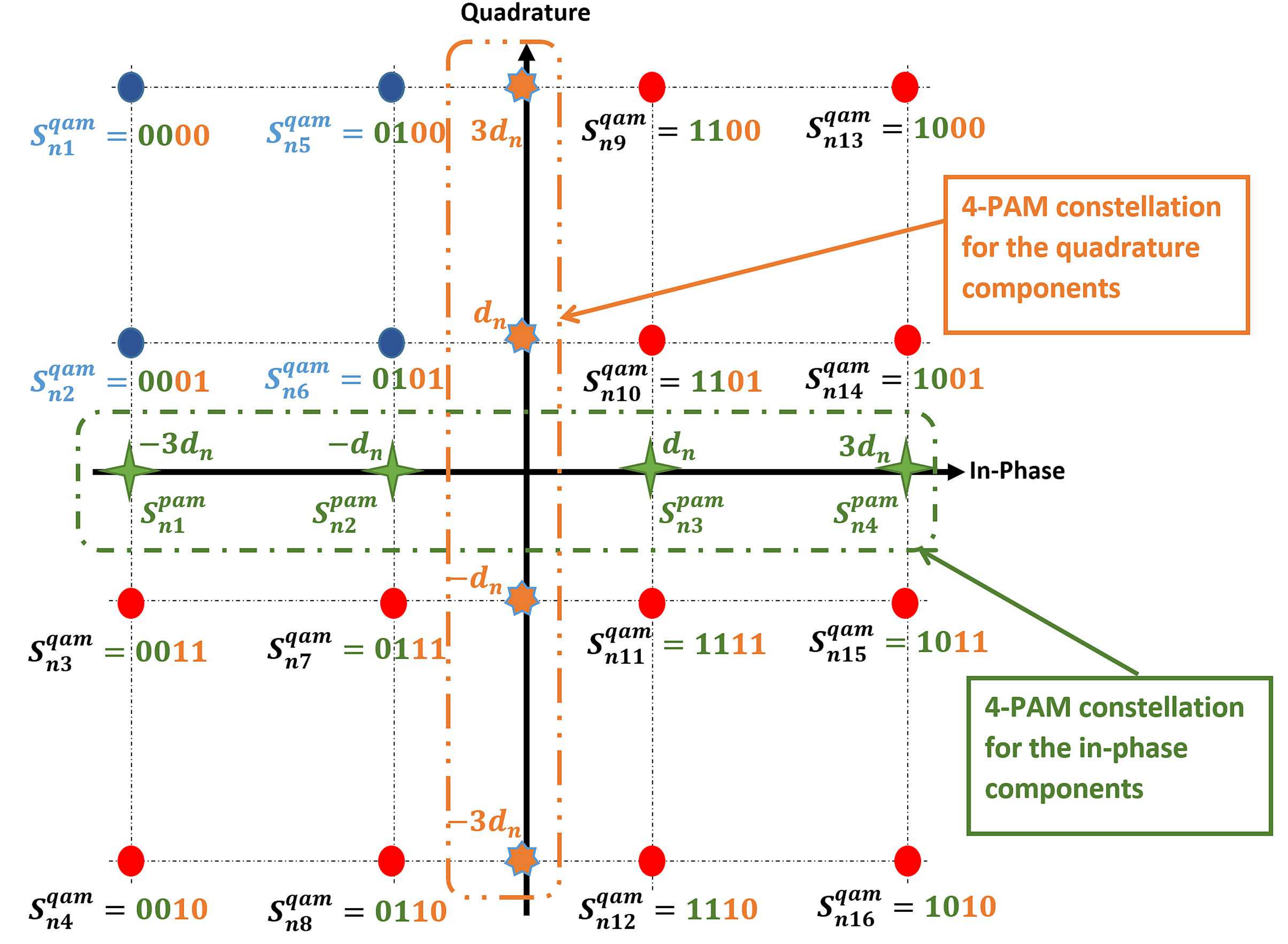}%{fig1.png}
    \caption{The constellation diagram of $16$-QAM modulation.}
    \label{fig1}
\end{figure}

The rest of this paper is organized as follows. The system and channel models are described in Section II. In Section III, we derive an upper bound of BER for multi-user uplink NOMA with adaptive $\mathcal{M}$-QAM. Simulation results are provided in Section IV. Finally, Section V concludes the paper.

\section{System Model}
We consider an uplink-NOMA scenario where one base station equipped with $L$ antennas supports $N$ active single-antenna users. All users share the same channel resources using their own powers $P_{n}$, $n=1, \cdots,N$.
Therefore, the received signal vector $\mathbf{y}\in\mathbb{C}^{L\times1}$ at the BS can be written as 
\begin{equation}\label{eq:1}
\mathbf{y}=\sum_{{n}=1}^{N} \mathbf{g}_n \sqrt{P_{n}} x_n +\mathbf{w},   \end{equation} 
where $\mathbf{g}_n\in\mathbb{C}^{L\times1}$  denotes the channel gain vector between the $n^{th}$ user and the BS. The components of the vector $\mathbf{g}_n$ are independent and identically distributed (i.i.d.) and follow $\mathbf{g}_n \sim \mathcal{CN}(0,\sigma_n^2\mathbf{I}_L)$, where  $\mathbf{I}_L$ denotes the $L \times L$ identity matrix. The vector $\mathbf{w}\in\mathbb{C}^{L\times1}$is the additive white Gaussian noise vector whose elements follow $\mathbf{w} \sim \mathcal{CN}(0,N_0\mathbf{I}_L)$. Moreover, it is assumed that the channel gains and the received signal powers are correctly measured at the BS, such that
 $ \left\|\mathbf{h}_1\right\|> \left\|\mathbf{h}_2\right\|>  \cdots >\left\|\mathbf{h}_N\right\|$, where $\mathbf{h}_n=\sqrt{P_{n}}\mathbf{g}_n$, $n=1,\cdots\,N$, and the symbol $\left\|.\right\|$ denotes the Frobenius norm. For each $n^{th}$ user (U$_n$), the information bits are modulated as complex symbol $x_n\in \mathbb{C}$, $n=1, \cdots,N$, where it takes its value from the alphabet $\mathbf{\chi}_{n}^{qam}=\left[ s_{n1}^{qam},s_{n2}^{qam},\cdots,s_{n\mathcal{M}_N}^{qam}\right]^T$, and the symbol $s_{ni_n}^{qam}$ is the $i_n^{th}$ constellation point in $\mathcal{M}_n$-ary modulation order of the $n^{th}$ user. In this letter, we consider that all users are served with an adaptive modulation (modulation order is determined according to the CQI as given in standards \cite{3gpp1630}) that uses a Gray-coded square $\mathcal{M}_n$-QAM, where $\mathcal{M}_n\geq4$ is power of 2. The amplitudes of the in-phase and the quadrature components of the QAM symbol are selected over the set of $\left\{\pm d_n, \pm 3d_n,\cdots,\pm \left( \sqrt{\mathcal{M}_n}-1\right)d_n \right\}$, where $d_n$ can be computed using the bit energy $E_b$ as
 \begin{equation} \label{eq:2}
d_n = \sqrt{\frac{3 E_b  \log_2{\mathcal{M}_n}}{2\left({\mathcal{M}_n}-1\right)}}. 
\end{equation}
In order to reliably recover the data of all users at the receiver side (BS), we propose to jointly estimate the $\mathcal{M}_n$-QAM symbols of the $N$ users with a maximum-likelihood detector (JMLD), as  \begin{equation} \label{eq:3}
\lbrack\Hat{x}_1,\Hat{x}_2,\cdots,\Hat{x}_N\rbrack= \underset{s_{ni_n}^{qam}}{\mathrm{argmin}}\Big \|\mathbf{y}-\sum\nolimits_{{n}=1}^{N} \mathbf{h}_n s_{ni_n}^{qam}\Big \|^2.
\end{equation} 
The power ranking of users $\left\|\mathbf{h}_1\right\|> \left\|\mathbf{h}_2\right\|>\cdots >\left\|\mathbf{h}_N\right\|$, has no effect on the performance of MRC-JMLD since the detector performs a search over all the possible combinations $\lbrack \mathbf{h}_1 s_{1i_{1}}^{qam},\mathbf{h}_2 s_{2i_{2}}^{qam},\cdots,\mathbf{h}_N s_{Ni_{N}}^{qam} \rbrack$. This explains the power of the joint detection to remove the error floor caused by the iterative detection 
(sequentially use the ML detector to detect each signal after removing the signals detected from the previous step) 
like MRC-SICD, which suffers from error propagation in bit error rate (BER) performance.

\section{BER Upper Bound Analysis}
 
Since an $\mathcal{M}_n$-QAM square modulation is a two-dimensional generalization of the $\mathcal{I}_n$-PAM modulation (see Fig. 1 for $16$-QAM representation), where $\mathcal{I}_n=\sqrt{\mathcal{M}_n}$, we first analyze the performance of MRC-JMLD for the case of one-dimension PAM modulation, then, the result is generalized for the case of bi-dimensional QAM modulation.

\subsection{Upper bound expression for $\mathcal{I}_n$-ary PAM}
\begin{figure*}
\setlength\belowcaptionskip{-.3\baselineskip}
\centering
\subfloat[{Error pattern for User1 (U$_1$)}]{\includegraphics[width=0.66\columnwidth]{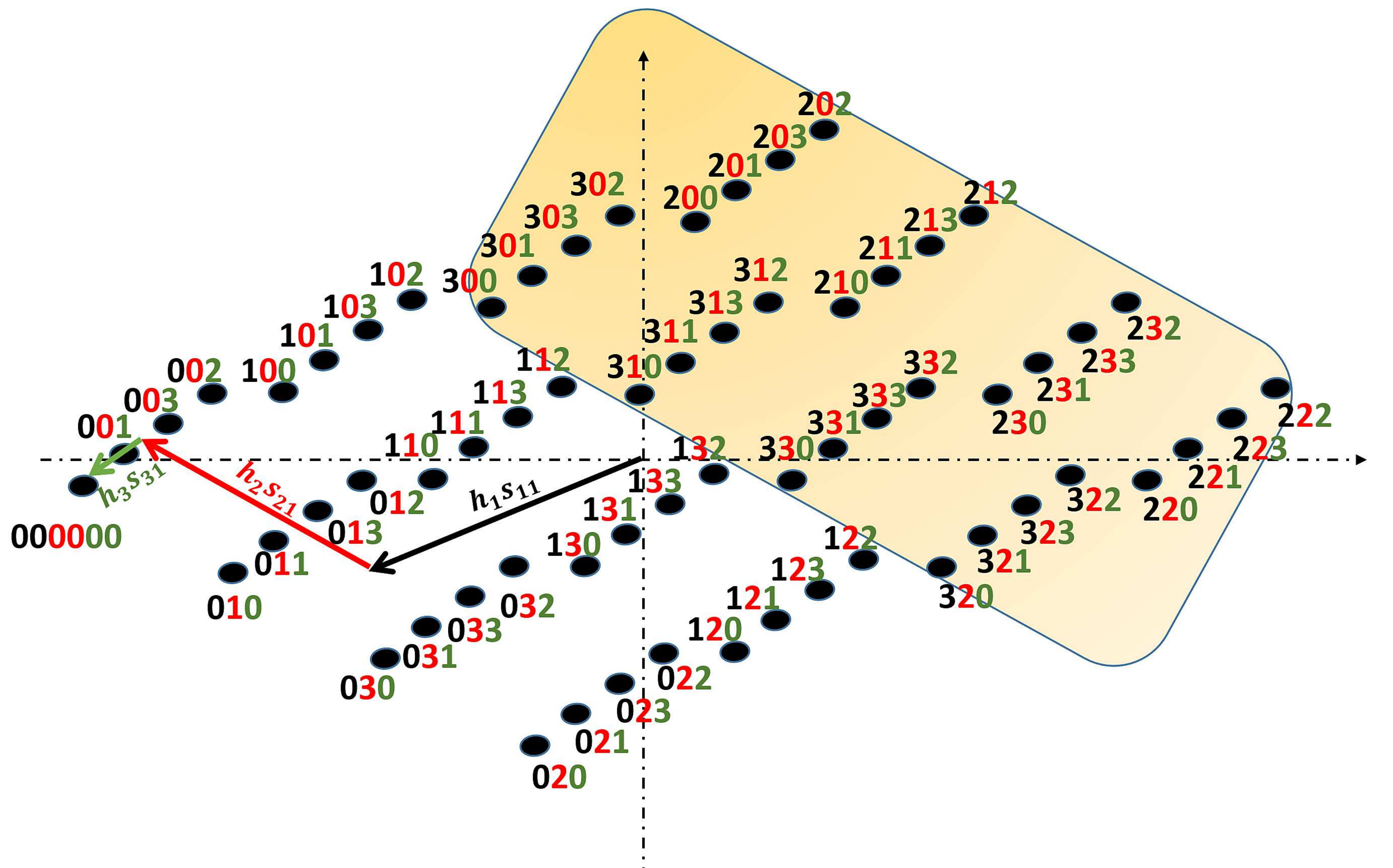}%{fig2a.png}
\label{img1:const1}}
\subfloat[{Error pattern for User2 (U$_2$)}]{\includegraphics[width=0.66\columnwidth]{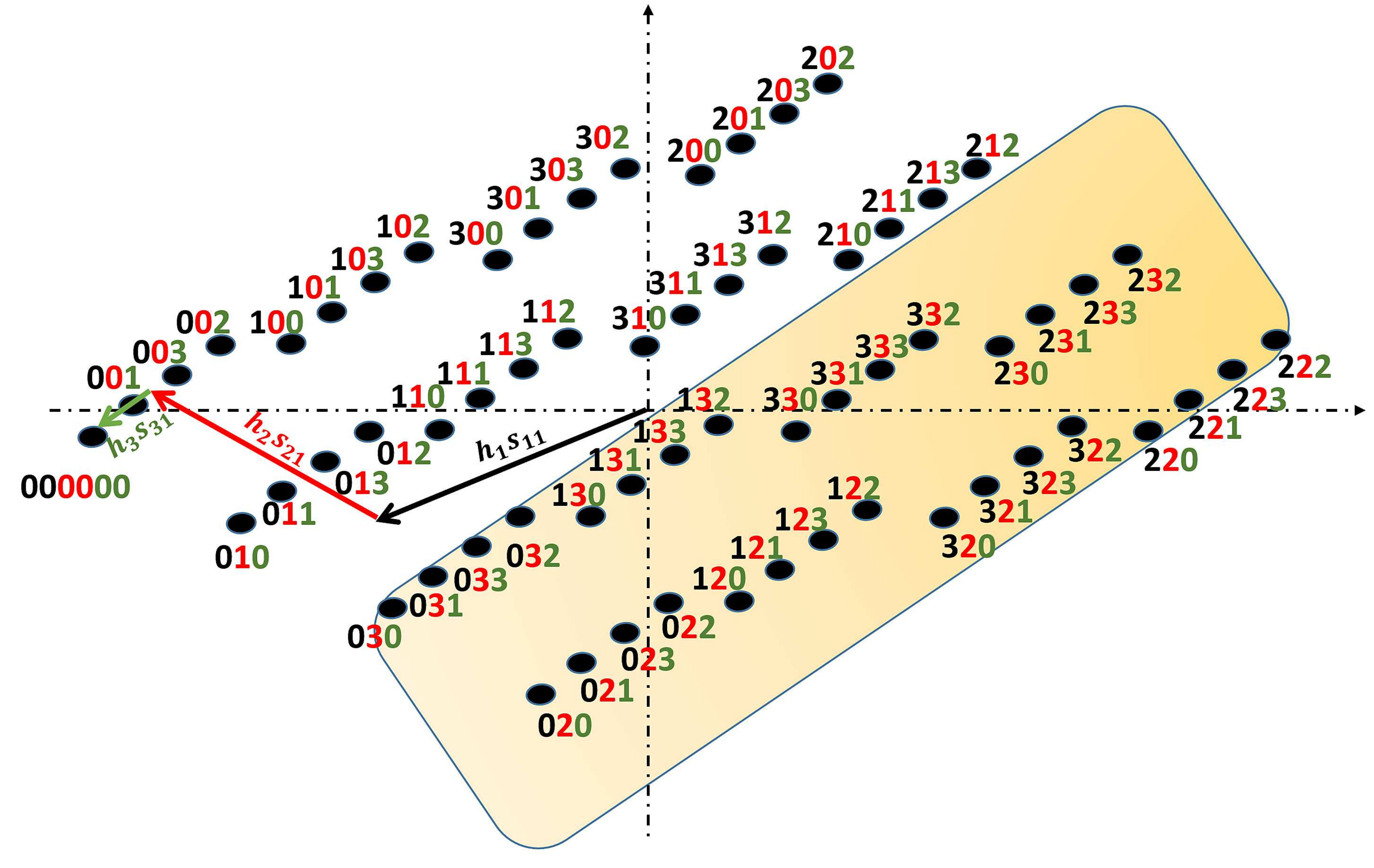}%{fig2b.png}
\label{img2:const2}}
\subfloat[{Error pattern for User3 (U$_3$)}]{\includegraphics[width=0.66\columnwidth]{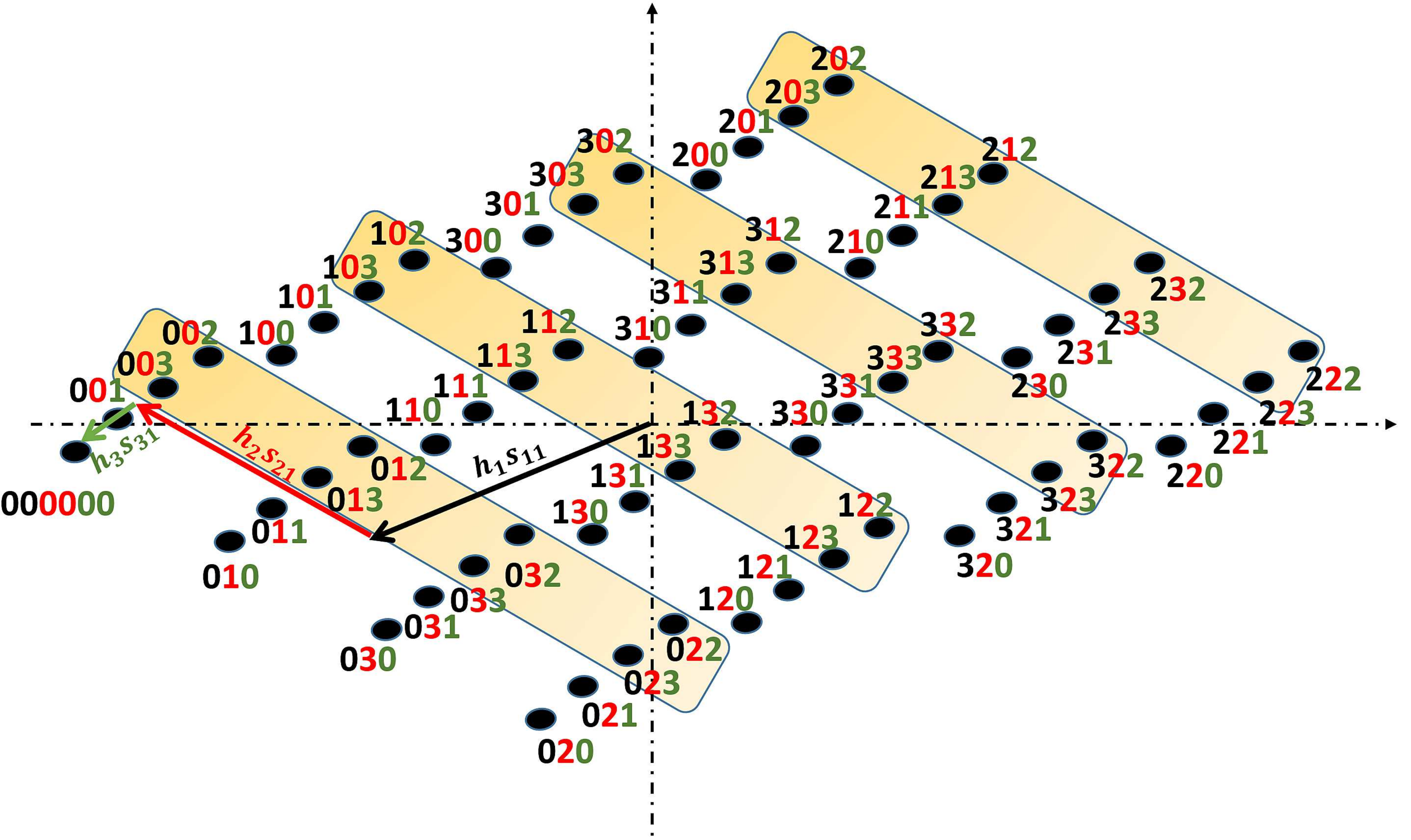}%{fig2c.png}
\label{img3:const3}}
\caption{{The constellation diagram of superimposed symbols of three users ($4$-PAM).}}
\label{constellations}
\end{figure*}

Without lost of generality, we begin our study by assuming a superposition of three $4$-PAM symbols (i.e., $N=3$ and $\mathcal{I}_n=4$) with equally energy per symbol $E_n=E_b\log_2{\mathcal{I}_n}=\frac{d_n^2 (\mathcal{I}_n^2-1)}{3}$, $n=1,2,\cdots,N$. Hence, the received signal at BS looks like one of the $\mathcal{I}_1\times\mathcal{I}_2\times\mathcal{I}_3=64$ constellation points shown in Fig. \ref{constellations}. Accordingly, each point in the constellation is represented by the binary sequence\footnote{For a clear illustration, we have adopted a quaternary representation of the symbols for the three users.} ${\{b_{11} b_{12} b_{21} b_{22} b_{31} b_{32}\}}$, where each bit is denoted by $b_{ni}$, $n=1, 2,3$, $i=1, 2$. 

Let us assume that all users send all zero bits which correspond to the binary sequence $\{00, 00, 00\}$. Owing to the symmetry  of the PAM constellation diagram (see the two leftmost bits in in-phase axis of Fig. \ref{fig1}), it is sufficient to limit the analysis on the left half plan that corresponds to $b_{n1}=0$, $n=1,2,3$. As the MRC-JMLD is based on the ML detector, the transmitted symbols are detected erroneously if the detector chooses one of the symbols located in the right half plane (the $\frac{\mathcal{I}_n}{2}=2$ symbols with first bit $b_{n1}=1$). By taking into consideration the separate detection of all users' symbols, the total number of erroneous superimposed symbols is equal to $\frac{4\times4\times4}{2}=32$ for each user (see the dashed areas in Figures \ref{img1:const1}, \ref{img2:const2} and \ref{img3:const3}). Consequently, by considering $N$ users with an adaptive modulation order of $\mathcal{I}_n$, the total number of symbols, which has an erroneous detection in the first bit, is equal to  $\frac{1}{2} \prod_{n=1}^{N} \mathcal{I}_n$. Due to the power ranking, the number of erroneous overlapped symbols is the same whatever the number of users with a difference in the error pattern.

To analyze the BER performance by using the union bound method\footnote{The presence of arbitrary multi-source interference where each user is independently affected by its path gain, alters the regularity of the decision region for NOMA-symbols; thence, the decision boundaries change according to the randomness of the amplitude and the phase with the modulation order. Accordingly, the derivation of an exact BER expression for more than two users is intractable. To surmount this problem, we accomplish an upper-bound analysis which gives clear insights for the BER curves in a practical SNR regime.} in the new signal-space diagram, we need to evaluate all the pairwise error probabilities (PEP) conditioned on the vectors $\textbf{h}_n$, $n=1,\cdots,N$, by regarding all combinations of symbols with the first bit $b_{n1}=0$. Thus, by exploiting the symmetry and the linearity of the $\mathcal{I}_n$-PAM constellation diagram, we can see that there are exactly $\frac{\mathcal{I}_n}{2}$ symbols (located on the left semi-straight of the $\mathcal{I}_n$-PAM constellation diagram) involved in the derivation of the PEPs. Therefore, for the same $n^{th}$ user within a NOMA symbol including the remaining superimposed symbols from the other users, the conditional PEP is related to the detecting of one of the symbols $s_{nj_n}^{pam}$, $j_n=\frac{\mathcal{I}_n}{2}+1,\cdots,\mathcal{I}_n$, as $b_{n1}= 1$ whereas $s_{ni_n}^{pam}$, $i_n=1,\cdots,\frac{\mathcal{I}_n}{2}$ is transferred as $b_{n1}= 0$. It is evaluated by 
\begin{equation} \label{Eq4} 
 Pr\left({s_{ni_{n}}^{pam}\to s_{nj_{n}}^{pam}} \mid {\mathbf{h}_1,\cdots,\mathbf{h}_N}\right)=
\rm{Q}\left(\frac{\parallel \mathfrak{Dt}_{i_nm}^{U_n^{pam}}\parallel
}{\sqrt{2  N_0}}\right),
 \end{equation}
where $\mathfrak{Dt}_{i_nm}^{U_n^{pam}}={2d_n\mathbf{h}}_{n}{\mathcal{D}}_{n,i_n}^{U_n^{pam}}(m)+$$\sum_{\substack{k=1 \\ k\neq n}}^{N}$$  {2d_k\mathbf{h}}_{k}{\mathcal{D}}_{k,i_k}^{U_n^{pam}}(m)$, and $m=1,\cdots,$ $\frac{1}{2}\prod_{i=1}^{N} \mathcal{I}_{i}$. For each U$_n$, the two real vectors  $2d_n\mathcal{D}_{n,i_n}^{U_n^{pam}}$ and  $2d_k\mathcal{D}_{k,i_k}^{U_n^{pam}}$ $\in\mathbb{R}^{\left(\frac{1}{2}\prod_{i=1}^{N} \mathcal{I}_{i}\right)\times1}$ denote the distances between one of the superimposed $\frac{\mathcal{I}_n}{2}$ symbols containing the symbol under test $s_{ni_{n}}^{pam}$ and its corresponding  $\frac{1}{2} \prod_{n=1}^{N} \mathcal{I}_n$ erroneous superimposed symbols that include $s_{nj_{n}}^{pam}$. Because of the superposition of different users' symbols, after being scaled by the factors $\mathbf{h}_n$, $n=1,\cdots,N$, the new signal-space diagram of the NOMA symbols is commonly a linear transformation of $\mathcal{I}_n$-PAM constellation points diagram. Hence, the linearity and the symmetry properties of $\mathcal{I}_n$-PAM constellation are well kept. By virtue of these properties, only the contribution of $\frac{\mathcal{I}_k}{2}$ symbols of the remaining users are considered, where $k=1,\cdots,N$ and $k\neq n$. Consequently, we count for each user exactly $\sum_{k=1}^N\frac{\mathcal{I}_k}{2}$  distances vectors of dimension   $\frac{1}{2}\prod_{i=1}^{N} \mathcal{I}_{i}$. As can be seen from Fig. \ref{constellations}, the Gray-coded NOMA symbols at the receiver follow a regular pattern which can be exploited to calculate all the distances between the symbols; hence, the vectors $\mathcal{D}_{k,i_k}^{U_n^{pam}}$ are given by \cite{semira2021_wcl} 
  \begin{equation} \label{Eq5} 
     \mathcal{D}_{k,i_k}^{U_n^{pam}}=
     \begin{cases}
      \mathbbm{1}_{(\prod_{i=1}^{n-1} \mathcal{I}_{i},1)} \otimes \mathfrak{E}_{n,i_{n}}^{U_n^{pam}}\otimes \mathbbm{1}_{(\prod_{i=n+1}^{N}\mathcal{I}_{i},1)}, \; k=n,\\
      \mathbbm{1}_{(\prod_{i=1}^{k-1} \mathcal{I}_{i},1)} \otimes \mathfrak{d}_{k,i_k}^{U_n^{pam}}\otimes \mathbbm{1}_{(\frac{1}{2} \prod_{i=k+1}^{N}\mathcal{I}_{i},1)} , \; k < n,\\
      \mathbbm{1}_{(\frac{1}{2} \prod_{i=1}^{k-1} \mathcal{I}_{i},1)} \otimes \mathfrak{d}_{k,i_k}^{U_n^{pam}}\otimes \mathbbm{1}_{(\prod_{i=k+1}^{N}\mathcal{I}_{i},1)}, \; k > n,
    \end{cases} 
 \end{equation}
where $\mathbbm{1}_{(\prod_{i=1}^{k-1} \mathcal{I}_{i},1)}$ is the all ones vector of dimension ${\prod_{i=1}^{k-1} \mathcal{I}_{i}}\times1$ and $\otimes$ denotes the Kronecker product. The vectors $\mathfrak{E}_{n,i_{n}}^{U_n^{pam}}=\frac{1}{2d_n}[s_{ni_n}^{pam}-s_{n\frac{\mathcal{I}_{n}}{2}+1}^{pam},\cdots,s_{ni_n}^{pam}-s_{n\mathcal{I}_{n}}^{pam}]^T$, $i_{n}=1,\cdots,\frac{\mathcal{I}_{n}}{2}$, and $\mathfrak{d}_{k,i_k}^{U_n^{pam}}=\frac{1}{2d_k}[s_{ni_k}^{pam}-s_{n1}^{pam},\cdots,s_{ni_k}^{pam}-s_{n\mathcal{I}_{k}}^{pam}]^T$, $i_{k}=1,\cdots,\frac{\mathcal{I}_{k}}{2}$, represent the distances between the different symbols based on their position within the constellation diagram, where their elements are calculated by
  \begin{equation} \label{Eq6} 
         \mathfrak{E}_{n,i_{n}}^{U_n^{pam}}= |i_{n}-(j_{n}+\frac{\mathcal{I}_{n}}{2})|,\;i_{n}=1,\cdots,\frac{\mathcal{I}_{n}}{2},\; j_{n}=1,\cdots,\frac{\mathcal{I}_{n}}{2},
   \end{equation}
and
\begin{equation} \label{Eq7} 
   % \begin{aligned}
   \mathfrak{d}_{k,i_{k}}^{U_n^{pam}}=|i_{k}-j_{k}|,\;
   i_{k}=1,\cdots,\frac{\mathcal{I}_{k}}{2},\; j_{k}=1,\cdots,\mathcal{I}_{k}.
   \end{equation}
   
At this stage, we have found the PEP by considering only the first bit $b_{n1}$ for each symbol $s_{n{i_n}}$, $i_n=1,\cdots,\frac{\mathcal{I}_n}{2}$. Thus, for the rest of the bits $b_{n{\log_2\mathcal{I}_n}}$ and by exploiting the symmetry of $\mathcal{I}_n$-PAM constellation, along with adaptation of equation \cite[Eq. (15)]{Jianhua99} for the case of PAM modulation in the new constellation diagram, we can obtain as
  \begin{equation} \label{Eq8} 
 P_{e,\mathcal{I}_n\textrm{-}pam}\left(e\mid {\mathbf{h}_1,\cdots,\mathbf{h}_N}\right)=\frac{(\mathcal{I}_n-1)}{\log_2\mathcal{I}_n}P_Q^{U_n^{pam}}.
   \end{equation}
Given that all symbols are equiprobable, the probability $P_Q^{U_n^{pam}}$ with respect to the center of the new constellation diagram is expressed as
  \begin{equation} \label{Eq9} 
P_Q^{U_n^{pam}} =\prod_{k=1}^{N}\frac{2}{\mathcal{I}_k}\Sum_{{i_n}=1}^{\frac{\mathcal{I}_n}{2}}\underbrace{\Sum_{{i_1}=1}^{\frac{\mathcal{I}_1}{2}}\cdots\Sum_{{i_N}=1}^{\frac{\mathcal{I}_N}{2}}}_{\text{N-1 sums}}\rm{Q}\left(\frac{\parallel \mathfrak{Dt}_{i_nm}^{U_n^{pam}}\parallel
}{\sqrt{2  N_0}}\right).
   \end{equation}
Substituting (\ref{Eq9}) into (\ref{Eq8}) and computing the upper bound for the probability of union of all events, we get as a result
\begin{multline} \label{Eq10} 
P_{e,\mathcal{I}_n\textrm{-}pam}\left(e\mid {\mathbf{h}_1,\cdots,\mathbf{h}_N}\right)\leq \frac{(\mathcal{I}_n-1)}{\log_2\mathcal{I}_n}\prod_{k=1}^{N}\frac{2}{\mathcal{I}_k}\\\Sum_{{i_n}=1}^{\frac{\mathcal{I}_n}{2}}\underbrace{\Sum_{{i_1}=1}^{\frac{\mathcal{I}_1}{2}}\cdots\Sum_{{i_N}=1}^{\frac{\mathcal{I}_N}{2}}}_{\text{N-1 sums}}\sum_{{m}=1}^{\frac{1}{2}\prod_{i=1}^{N} \mathcal{I}_{i}} \rm{Q}\left(\sqrt{\Omega_{i_nm}^{pam}}\right),
\end{multline}
where $\Omega_{i_nm}^{pam}=\frac{{\parallel \mathfrak{Dt}_{i_nm}^{U_n^{pam}}\parallel}^2}{2  N_0}$ is the received superimposed symbol signal-to-noise ratios (\textsf{SNR}$_{i_{n}}$) for diversity paths. Taking account of hypothesis $\mathbf{g}_n \sim \mathcal{CN}(0,\sigma_n^2\mathbf{I}_L)$, therefore, $\Omega_{i_nm}^{pam}\sim\mathcal{CN}(0,4d_{n}^{2}P_{n}\sigma_{n}^{2}|{\mathcal{D}}_{n,i_n}^{U_n^{pam}}(m)|^{2}+\sum_{\substack{k=1 \\ k\neq n}}^{N}{4d_{k}^{2}P_{k}\sigma_{k}^{2}|{\mathcal{D}}_{k,i_k}^{U_n^{pam}}(m)|^{2}})$. By considering the maximum ratio combining of $L$ uncorrelated fading paths, ${\Omega_{i_nm}^{pam}}$ follows a chi-square PDF with $2L$ degrees of freedom \cite{Simon2005} (sum of $L$ i.i.d. Rayleigh fading branches) such that %\cite{Simon2005}
\begin{equation} \label{Eq11}
 P_{\Omega_{i_nm}^{pam}}\left(\zeta\right)=\frac{1}{\left(L-1\right)!}\frac{\zeta^{L-1}}{{\left(\Gamma_{i_nm}^{U_n^{pam}}\right)}^L} \exp{\left(\frac{-\zeta}{\Gamma_{i_nm}^{U_n^{pam}}}\right)},
 \end{equation}
where \small$\Gamma_{i_nm}^{U_n^{pam}}=\gamma_{n}^{pam}|{\mathcal{D}}_{n,i_n}^{U_n^{pam}}(m)|^{2}+\sum_{\substack{k=1 \\ k\neq n}}^{N}\gamma_{k}^{pam}|{\mathcal{D}}_{k,i_k}^{U_n^{pam}}(m)|^{2}$ \normalsize is the average \textsf{SNR}$_{i_{n}}^{pam}$ per superimposed symbol linked to $s_{i_{n}}^{pam}$ and $\gamma_{n}^{pam}=\frac{6E_b\log_2{\mathcal{I}_n}}{\left({\mathcal{I}_n}^2-1\right)}\frac{P_n\sigma_n^2}{N_0}$. By averaging the upper bound of (\ref{Eq10}) over the distribution of $\Omega_{i_nm}^{U_n^{pam}}$, we find
 % \begin{equation} 
  \begin{multline} \label{Eq12} 
  P_{e,\mathcal{I}_n\textrm{-}pam}\leq \frac{(\mathcal{I}_n-1)}{\log_2\mathcal{I}_n}\prod_{k=1}^{N}\frac{2}{\mathcal{I}_k}\Sum_{{i_n}=1}^{\frac{\mathcal{I}_n}{2}}\\
 \underbrace{\Sum_{{i_1}=1}^{\frac{\mathcal{I}_1}{2}}\cdots\Sum_{{i_N}=1}^{\frac{\mathcal{I}_N}{2}}}_{\text{N-1 sums}}
\displaystyle \sum_{{m}=1}^{\frac{1}{2}\prod_{i=1}^{N} \mathcal{I}_{i}} \int_{0}^\infty \rm{Q}\left(\sqrt{\zeta}\right)P_{\Omega_{i_nm}^{pam}}\left(\zeta\right)d\zeta.
\end{multline}

Furthermore, utilising \cite[Eq. (7)]{Yeom2019} to evaluate the integral of (\ref{Eq12}) yields the closed form expression of (\ref{Eq13}). Note that this expression represents a generic form of the average BER upper bound for the $n^{th}$ user in a competitive environment with $N$ users using adaptive $\mathcal{I}_n$-PAM modulation. 
  \begin{figure*}
     \begin{equation} \label{Eq13}
 P_{e,\mathcal{I}_n\textrm{-}pam}^{U_n}\leq
\frac{(\mathcal{I}_n-1)}{2\log_2\mathcal{I}_n}\prod_{k=1}^{N}\frac{2}{\mathcal{I}_k}\Sum_{{i_n}=1}^{\frac{\mathcal{I}_n}{2}}\underbrace{\Sum_{{i_1}=1}^{\frac{\mathcal{I}_1}{2}}\cdots\Sum_{{i_N}=1}^{\frac{\mathcal{I}_N}{2}}}_{\text{N-1 sums}}
\sum_{{m}=1}^{\frac{1}{2}\prod_{i=1}^{N} \mathcal{I}_{i}} \left[ 1-\sum\nolimits_{l=0}^{L-1} {\binom{2l}{l}} {\left({1+2/\Gamma_{i_nm}^{U_n^{pam}}}\right)^{-\frac{1}{2}}} {\left(2\Gamma_{i_nm}^{U_n^{pam}}+4\right)^{-l}} \right].
\end{equation}
%\hrulefill
\end{figure*}
\subsection{Upper bound expression for $\mathcal{M}_n$-ary square QAM}
As stated in the beginning of Sec. III, the square $\mathcal{M}_n$-QAM modulation can be seen as a combination of two independent PAM modulations of $\sqrt{\mathcal{M}_n}$ size; consequently, the analysis of the performance of NOMA signals with the $\mathcal{I}_n$-PAM modulation can be generalized to the two-dimensional square Gray-coded $\mathcal{M}_n$-QAM modulation. Similar to the NOMA-PAM modulation, the new signal space is simply a rotation of the different signals spaces of the superimposed QAM symbols scaled by the coefficients of the channel. Compared to the previous analysis and based on the geometry of the signal space diagram in Fig. \ref{fig1}, if we consider only the first bit for any user, an error of detection arises when the BS chooses one of the symbols located in the right half plane of the constellation diagram. Therefore, we count exactly  $\frac{1}{2}\prod_{i=1}^{N}\mathcal{M}_{i}$  erroneous symbol for each user. Moreover, due to the symmetry of the QAM constellation, the distances specified between the points of the left upper quadrant (first bit $b_{n1} = 0$) and those of the right quadrant (first bit $b_{n1} = 1$) appear to be equal in pairs in relation to those computed from the lower right quadrant. As a consequence, the PEP calculation only concerns the symbols in the upper left quadrant of the constellation diagram (the blue constellation points of Fig.\ref{fig1}). Thence, it suffices to take into account the contribution of the $\frac{\mathcal{M}_{n}}{4}$,$n=1,\cdots,N$, symbols to derive the PEPs. Following the same reasoning for the superimposed symbols from the remaining $N-1$ users, only the contribution of $\frac{\mathcal{M}_{k}}{4}$ symbols is needed.
\begin{figure*}
\setlength\belowcaptionskip{-.5\baselineskip}
     \begin{equation} \label{Eq14}
 P_{e,\mathcal{M}_n\textrm{-}qam}^{U_n}\leq
\frac{(\sqrt{\mathcal{M}_n}-1)}{\log_2\mathcal{M}_n}\prod_{k=1}^{N}\frac{4}{\mathcal{M}_k}\Sum_{{i_n}=1}^{\frac{\mathcal{M}_n}{4}}\underbrace{\Sum_{{i_1}=1}^{\frac{\mathcal{M}_1}{4}}\cdots\Sum_{{i_N}=1}^{\frac{\mathcal{M}_N}{4}}}_{\text{N-1 sums}}
\sum_{{m}=1}^{\frac{1}{2}\prod_{i=1}^{N} \mathcal{M}_{i}} \left[ 1-\sum\nolimits_{l=0}^{L-1} {\binom{2l}{l}} {\left({1+2/\Gamma_{i_nm}^{U_n^{qam}}}\right)^{-\frac{1}{2}}} {\left(2\Gamma_{i_nm}^{U_n^{qam}}+4\right)^{-l}} \right].
\end{equation}
\hrulefill
\end{figure*}

Now, if we assume that the symbols are equally likely, one can do analogous procedures like the previous subsection to obtain the expression (\ref{Eq14}) (see the
top of the page), where $\Gamma_{i_nm}^{U_n^{qam}}= \gamma_{n}^{qam}|{\mathcal{D}}_{n,i_n}^{U_n^{qam}}(m)|^{2}+\sum_{\substack{k=1 \\ k\neq n}}^{N}\gamma_{k}^{qam}|{\mathcal{D}}_{k,i_k}^{U_n^{qam}}(m)|^{2}$  is the average \textsf{SNR}$_{i_{n}}^{qam}$ per superimposed symbol associated to $s_{i_{n}}^{qam}$ and $\gamma_{n}^{qam}=\frac{3E_b\log_2{\mathcal{M}_n} }{\left({\mathcal{M}_n}-1\right)}\frac{P_n\sigma_n^2}{N_0}$. Following the pattern of the structured Gray coded NOMA symbols, the two complex vectors $\mathcal{D}_{n,i_n}^{U_n^{qam}}$  and  $\mathcal{D}_{k,i_k}^{U_n^{qam}}$ of dimension $\frac{1}{2}\prod_{i=1}^{N}\mathcal{M}_{i}$ are defined as  
 \begin{equation} \label{Eq15} 
 \mathcal{D}_{k,i_k}^{U_n^{qam}}=
  \begin{cases}
    \mathbbm{1}_{(\prod_{i=1}^{n-1} \mathcal{M}_{i},1)} \otimes \mathfrak{E}_{n,i_{n}}^{U_n^{qam}}\otimes \mathbbm{1}_{(\prod_{i=n+1}^{N}\mathcal{M}_{i},1)}, k=n,\\
   \mathbbm{1}_{(\prod_{i=1}^{k-1} \mathcal{M}_{i},1)} \otimes \mathfrak{d}_{k,i_k}^{U_n^{qam}}\otimes \mathbbm{1}_{(\frac{1}{2} \prod_{i=k+1}^{N}\mathcal{M}_{i},1)}, k < n,\\
  \mathbbm{1}_{(\frac{1}{2} \prod_{i=1}^{k-1} \mathcal{M}_{i},1)} \otimes \mathfrak{d}_{k,i_k}^{U_n^{qam}}\otimes \mathbbm{1}_{(\prod_{i=k+1}^{N}\mathcal{M}_{i},1)}, k > n,
   \end{cases}
 \end{equation}
where the vector $\mathfrak{E}_{n,i_{n}}^{U_n^{qam}} \in \mathbb{C}^{\frac{\mathcal{M}_n}{2}\times 1}$, $i_{n}=1,\cdots,\frac{\mathcal{M}_{n}}{4}$, denotes the distances between one of the symbols in the upper left quadrant and the symbols $\frac{M_n}{2}$ in the right plan for the $n^{th}$ user, and the second vector which describes the distances between one of the symbols in the upper left quadrant and all the symbols for the remaining $N-1$ users is denoted by $\mathfrak{d}_{k,i_k}^{U_n^{qam}} \in \mathbb{C}^{{M}_k\times1}$, $i_{k}=1,\cdots,\frac{\mathcal{M}_{k}}{4}$, where $k\neq n$. From the relation between the PAM modulation and the QAM modulation, the last two vectors can be arranged in matrix form $\mathfrak{E}_{n}^{U_n^{qam}} \in \mathbb{C}^{\frac{\mathcal{M}_n}{2}\times \frac{\mathcal{M}_n}{4}}$ and $\mathfrak{d}_{k}^{U_n^{qam}} \in \mathbb{C}^{\mathcal{M}_k\times\frac{\mathcal{M}_k}{4}}$, as follows.
  \begin{equation} \label{Eq16} 
    \mathfrak{E}_{n}^{U_n^{qam}}= \mathfrak{E}_{n}^{U_n^{pam}}\otimes\mathbbm{1}_{(\mathcal{I}_{n},\frac{\mathcal{I}_{n}}{2})}+\sqrt{-1} \times  \mathbbm{1}_{(\frac{\mathcal{I}_{n}}{2},\frac{\mathcal{I}_{n}}{2})}\otimes \mathfrak{d}_{n}^{U_n^{pam}},
  \end{equation}
and  
  \begin{equation} \label{Eq17} 
    \mathfrak{d}_{k}^{U_n^{qam}}= \mathfrak{d}_{k}^{U_n^{pam}}\otimes\mathbbm{1}_{(\mathcal{I}_{n},\frac{\mathcal{I}_{n}}{2})}+\sqrt{-1} \times  \mathbbm{1}_{(\mathcal{I}_{n},\frac{\mathcal{I}_{n}}{2})}\otimes \mathfrak{d}_{k}^{U_n^{pam}},
  \end{equation}
where the columns of $\mathfrak{E}_{n}^{U_n^{pam}} \in \mathbb{N}^{\frac{\mathcal{I}_n}{2}\times\frac{\mathcal{I}_n}{2}}$ and $\mathfrak{d}_{n}^{U_n^{pam}}\in \mathbb{N}^{\mathcal{I}_n\times\frac{\mathcal{I}_n}{2}}$ are defined by (\ref{Eq6}) and (\ref{Eq7}), respectively. 

It should be noted that the expression (\ref{Eq14}) is the general formula of BER upper bound for an Uplink-NOMA access with $N$ users served with an adaptive $\mathcal{M}_n$-QAM modulation, where $\mathcal{M}_n \geq 4$. Taking into consideration the monotonicity behavior of the expressions (\ref{Eq13}) and  (\ref{Eq14}), one can infer that for large superimposed signal-to-noise ratio $\Gamma_{i_nm}^{U_n}$ the BER $\to 0$. This points out that the optimum detector MRC-JMLD is very efficient to eliminate the error floor. Instead, the iterative version of ML with interference cancellation, commonly used in the literature and nominated MRC-SICD, has a poor performance in a high SNR regime \cite{Kara2018d,Kara2020,Semira2021,semira2021_wcl}, and it completely fails to separate the users in this regime.
 
\section{Numerical Results}

In this section, computer simulations are presented to validate the theoretical analysis. Besides, the performance comparisons of the MRC-JMLD are also presented with the benchmark (i.e., MRC-SICD), which is the main detection algorithm in the literature. We set $L=4$ and to highlight the performance of the proposed scheme, we present three scenarios by varying the modulation order according to the CQI as given in Table 1. Thus, different receive powers are assigned to each user by changing the value of $P_n$, where we assume that $P_1=P_2=\cdots=P_N$, which reflects the practical applications in the real world where all users (e.g., IoT devices) have the same transmit power. In addition, this is the worst scenario in terms of interference where, for each user, the desired signal (own symbols) and the interference (other users' symbols) are conveyed with the same power. Even in this worst scenario, JML is capable of removing the error floor and it is obvious that the performance of the JML will be improved in less-interference scenarios.

\begin{table} 
\centering
\caption{Simulation parameters}
 \begin{tabular} {|c||c|c|c|}      
   \hline
  & 
  $N$ & $\mathcal{M}_1, \ M_2 \dots,M_N$ & $\left\|\mathbf{g}_1\right\|^2,\left\|\mathbf{g}_2\right\|^2, \dots, \left\|\mathbf{g}_N\right\|^2 $ (dB) \\
      \hline
     Scenario I & $2$  &$ 256, 16 $& $0, -3$\\
    \hline
      Scenario II  & $3$  & $16, 16, 16$ &  $0, -3, -6$\\
     \hline 
   Scenario III  & $4$ &$256, 64, 16, 4$ &  $0, -3, -6, -9$\\
  \hline
\end{tabular}
\label{Tab:tb1}
\end{table}

In Fig. 3, we present performance evaluation for scenarios in Table I. Firstly, one can notice that the derived upper bound matches well with the simulations for all scenarios, and it is very tight in the high SNR regime (almost the same). This reveals the correctness of the derivations. Besides, regardless of the number of users, the MRC-JMLD outperforms the benchmark. Indeed, the superiority of the MRC-JMLD becomes exceptional with increasing number of users. This is explained by the error propagation from one step to another in the SICD, which creates an error floor. On the other hand, MRC-JMLD eliminates the error floor by performing a joint detection simultaneously. Besides, it can be observed from the curves that MRC-JML attains and guarantees the full diversity order for all users (i.e., $\lim_{\Gamma\rightarrow\infty}\sfrac{\log P_e}{\log\Gamma}$).

\begin{figure*}[htbp]
\setlength\belowcaptionskip{-.6\baselineskip}
\centering
\subfloat[{}]{\includegraphics[width=0.63\columnwidth]{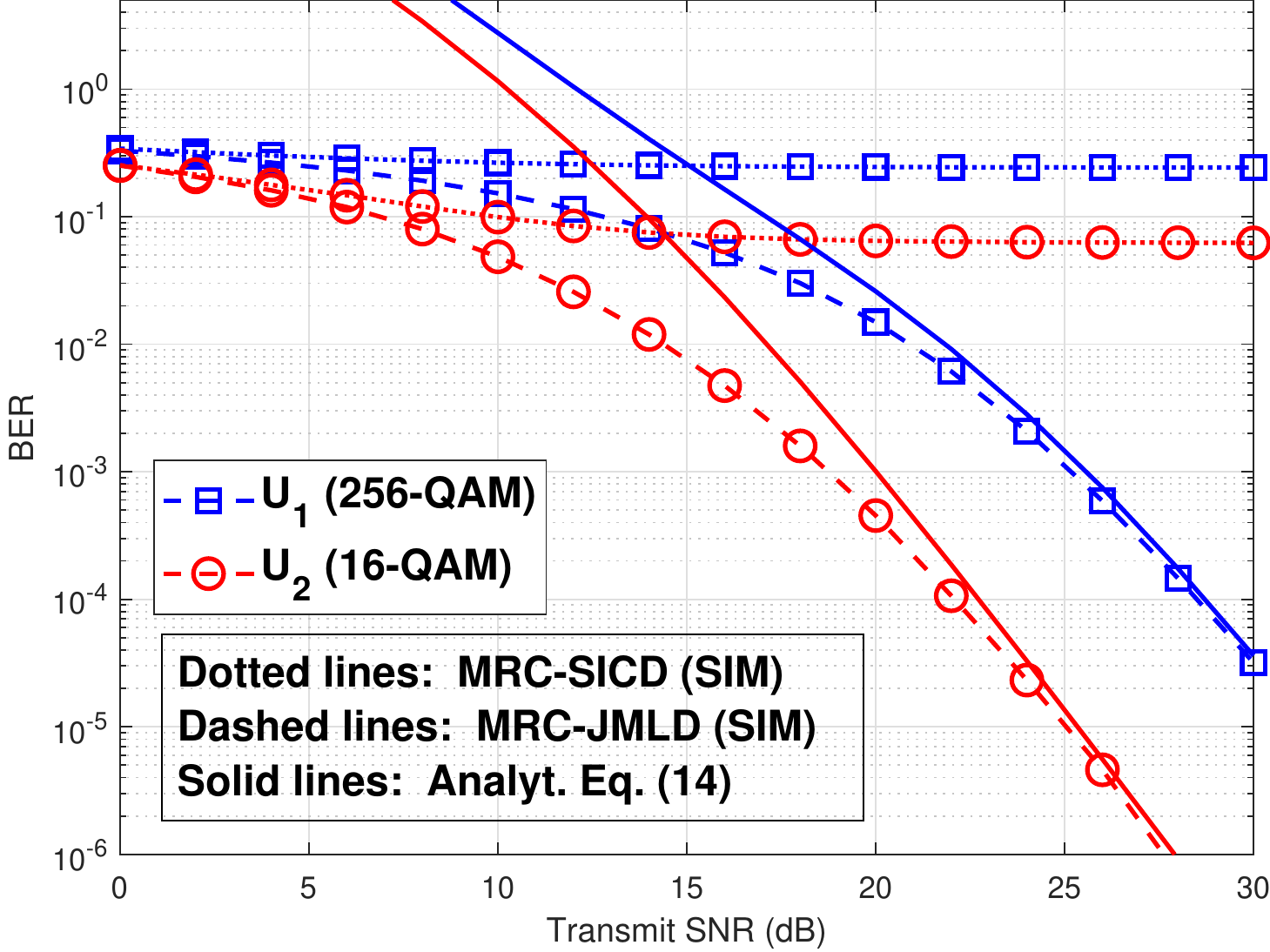}%{U64P_bis.eps}
\label{2U_ad}}
\subfloat[]{\includegraphics[width=0.63\columnwidth]{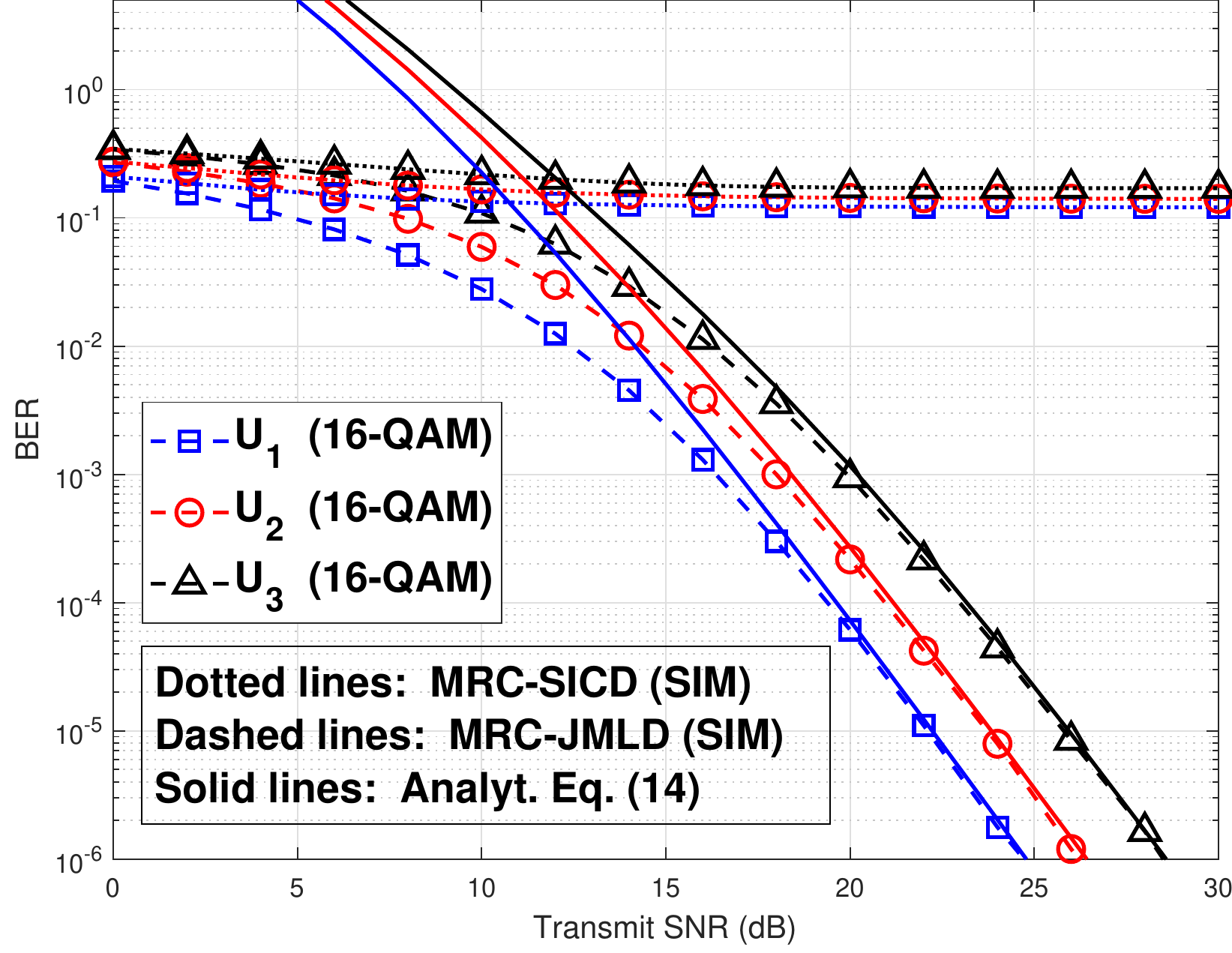}
\label{3U_16}}
\subfloat[{}]{\includegraphics[width=0.63\columnwidth]{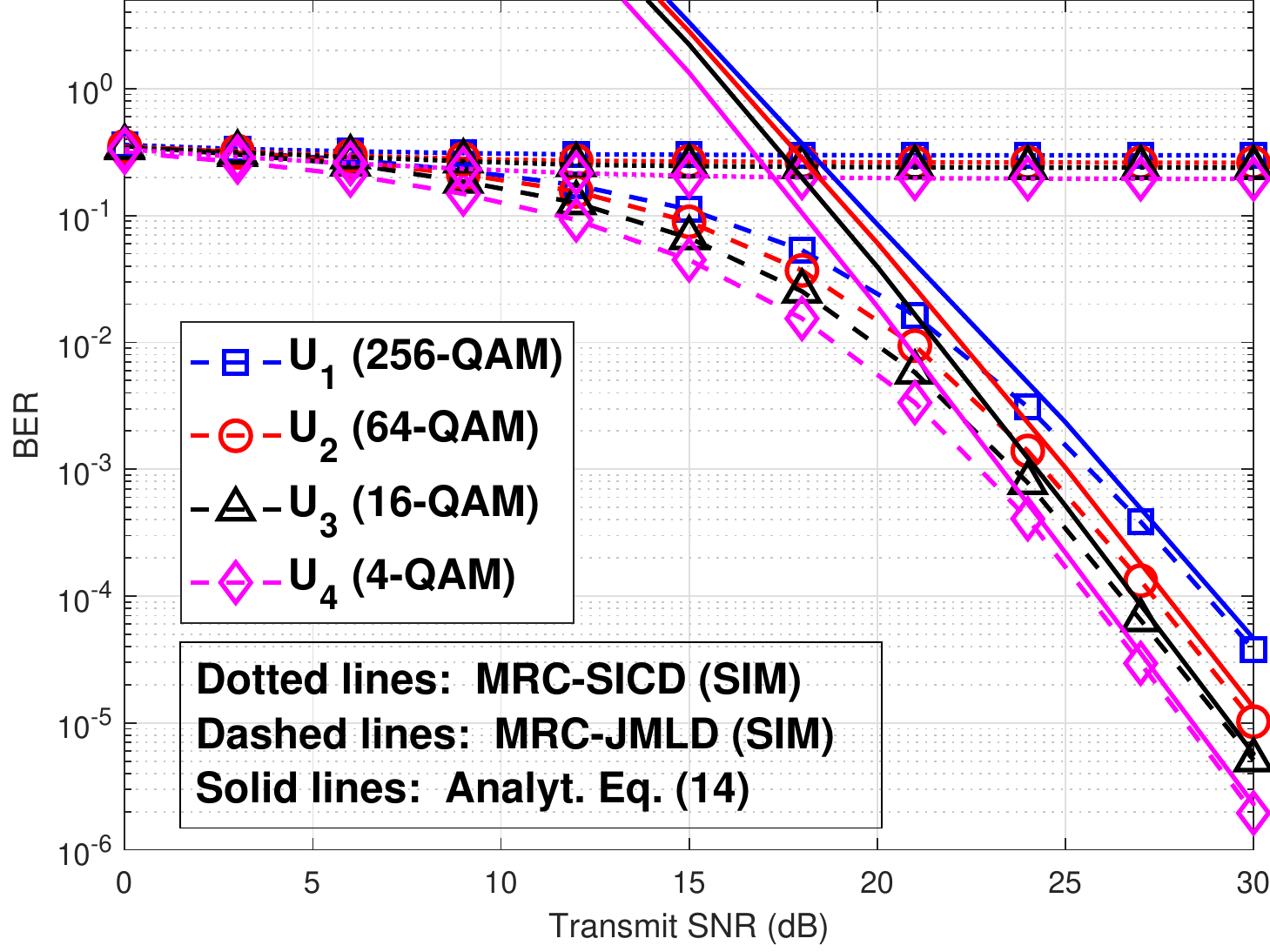}
\label{4U_ad}}
\caption{BER performance comparisons of JMLD and SICD in uplink NOMA with $L=4$ a) Scenario I b) Scenario II c) Scenario III}
\label{4D}
\end{figure*}

In Fig. 3.a, based on CQI, we consider $\mathcal{M}_1=256$ and $\mathcal{M}_2=16$; thus,  $\gamma_{1}^{qam} < \gamma_{2}^{qam}$, which explains the improvement of the performance of U$_2$ compared to U$_1$ despite the quality of its channel (i.e., a trade-off between the minimum euclidean distances $d_n$ with powers ($P_n$ and $\sigma_n^2$) and BER for each user).  For the second scenario, where we consider fixed modulation orders $\mathcal{M}_n=16$ in the case of three users ($N=3$), Fig. \ref{3U_16} shows that the BER of the users with the strongest gain channel is better than that of those with weak channels. As the minimum Euclidean distance is the same for all users, this clearly explains the influence of CQI on the performance of the system. In Fig. 3.c, the BER results are presented for different modulation orders ($256,64,16,4$) in the case of $N=4$ users. To achieve high channel capacity by increasing the data rate during good channel conditions, the modulation orders are adapted according to the CQI, where a higher order modulation is assigned to the best channel conditions, and the lower order is attributed to the poorer channel such as $\mathcal{M}_1=256$, $\mathcal{M}_2=64$, $\mathcal{M}_3=16$ and $\mathcal{M}_4=4$. We can observe the influence of the number of users on the performance of MRC-JMLD in comparison with the BER results of Fig. \ref{3U_16}, and this is related to the increase of the correlation between the data of users (i.e., increasing the multiaccess interference). As was anticipated, the modulation order has improved the performance of users with respect to their CQI, as a result of the expansion of the distance between symbols $d_n$ for the lower modulation order, where its influence is clearly visible through the parameters $\gamma_{1}^{qam} < \gamma_{2}^{qam}<\gamma_{3}^{qam} < \gamma_{4}^{qam}$.

\section{Conclusion}

This letter presents the error performance analysis of multi-user detection (MRC-JMLD) in uplink NOMA. We derive a tight upper bound BER expression of MRC-JMLD by using an adaptive $\mathcal{M}$-QAM determined by the CQI. Based on the extensive simulations, we reveal the superiority of the MRC-JMLD over MRC-SICD (the benchmark in the literature), where the MRC-JMLD removes the error floor and regardless of the number of users or the modulation order, it provides a full diversity order for all users. This letter reveals the potential of the JMLD for enabling massive connections by using NOMA (e.g., IoT applications) \cite{Yuan2021}. 
\bibliographystyle{IEEEtran}
\bibliography{Semira_WCL2022-0638_bib}
\end{document}